# All-dielectric free-electron-driven holographic light sources

Brendan P. Clarke[1], Behrad Gholipour[1, 2], Kevin F. MacDonald[1], and Nikolay I. Zheludev[1, 3]

[1] *Optoelectronics Research Centre & Centre for Photonic Metamaterials,  
University of Southampton, Southampton, SO17 1BJ, UK*

[2] *School of Chemistry, University of Southampton, Southampton, SO17 1BJ, UK*

[3] *Centre for Disruptive Photonic Technologies, School of Physical and Applied Sciences  
& The Photonics Institute, Nanyang Technological University, Singapore 637371*

It has recently been shown that holographically nanostructured surfaces can be employed to control the wavefront of (predominantly plasmonic) optical-frequency light emission generated by the injection of medium-energy electrons into a gold surface. Here we apply the concept to manipulation of the spatial distribution of transition radiation emission from high-refractive-index dielectric/semiconductor target materials, finding that concomitant incoherent luminescent emission at the same wavelength is unperturbed by holographic surface-relief structures, and thereby deriving a means of discriminating between the two emission components.

Imaging and spectroscopic analysis of cathodoluminescent emission – the light generated by the impact of free electrons on a material, are long-established techniques in electron microscopy, where they form part of the analytical toolkit for identifying dopants in semiconductors or mineral concentrations in geological surveys, and probing structural features such as fractures, stress regions and crystal interfaces.[1-5] In recent years, in particular following demonstrations of the fact that electron impact and can efficiently excite propagating and localized surface plasmons on metallic targets, which can subsequently couple to free-space light modes,[6, 7] a range of spatially-, temporally-, emission direction- and polarization-resolved hyperspectral electron-induced radiation emission (EIRE) imaging techniques that have been developed for the study of surface plasmon polariton propagation, mapping of plasmonic nanoparticle modes, and the identification of structural phase states.[8-17]

The availability of these characterization techniques and parallel advances in nanofabrication technologies have led to growing interest in frequency-tuneable free-electron-driven nanoscale light sources: A variety of plasmonic nanoantennas,[18, 19] metal-dielectric 'light-well' undulators,[20] Smith-Purcell gratings,[21, 22] plasmonic and photonic crystals,[23, 24] and metasurface resonator ensembles[25] have been employed to couple medium-energy free-electron excitations (via proximity and impact interactions) to well-defined free-space light modes. Finally, it has recently been shown that holographic surface-relief plasmonic sources[26] can provide control, by design, over the wavelength and wavefront of light emission resulting from the point-injection of medium-energy electrons into a gold surface (Fig. 1).

There are several material-dependent mechanisms by which light may be generated as the result of electrons impacting a surface. An electron crossing the boundary between two different materials releases energy proportional to the Lorentz factor of the particle in the form of 'transition radiation' (TR)[27] with spectral and spatial distributions and an intensity determined by the difference between the relative permittivities (c.f. refractive indices) of the two materials. On metal surfaces such impacts also generate surface plasmon polaritons (SPPs) with a broad (again material-dependent) spectral distribution. For certain metals at certain frequencies and electron energies electron energy may couple more efficiently to SPPs than to TR, but the former can only contribute to free-space (far-field) light emission by scattering at surface defects or engineered decoupling structures (e.g. gratings). (In the event that electrons are travelling faster than the speed of light in the target medium, Cerenkov radiation will also be generated, but this mechanism is not relevant to the present study.) All of the above are coherent emission processes, whereby the excitation is near-instantaneous and light is emitted from effective 'point-source' regions that are small compared with the wavelength of light. Incoherent processes such as direct and indirect carrier recombination dominate the emission of many semiconductors and dielectrics. These occur over time, often decaying gradually as electrons scatter many times within a relatively large interaction volume beneath the surface of a material, and can be spectrally sensitive to factors including temperature, strain, dopants/impurities, lattice defects, a quantum/structural resonances.[1-5, 28-31] These coherent and incoherent EIRE mechanisms rarely manifest in isolation though and their contributions are not readily disentangled in measurements of electron-induced light emission.[32]

For example, in the case of the gold holographic light sources of Ref. 26, the 30 keV electron excitation is coupled to a combination of SPPs and TR with an expected photons-per-electron efficiency ratio[33] of approximately 3:2 at the experimental wavelength of 800 nm. It is seen that holographic metasurface structures can very effectively convert these divergent emissions emanating from the electron impact point into light beams with selected wavefronts, specifically directional plane waves and high-order optical vortex beams. However, the holographic design process does not distinguish between the TR and SPP components of the excitation - both are part of the same singular 'reference' electric field distribution, and measurements do not discern the relative efficiency with which the two components of emission are coupled to the desired output beam.

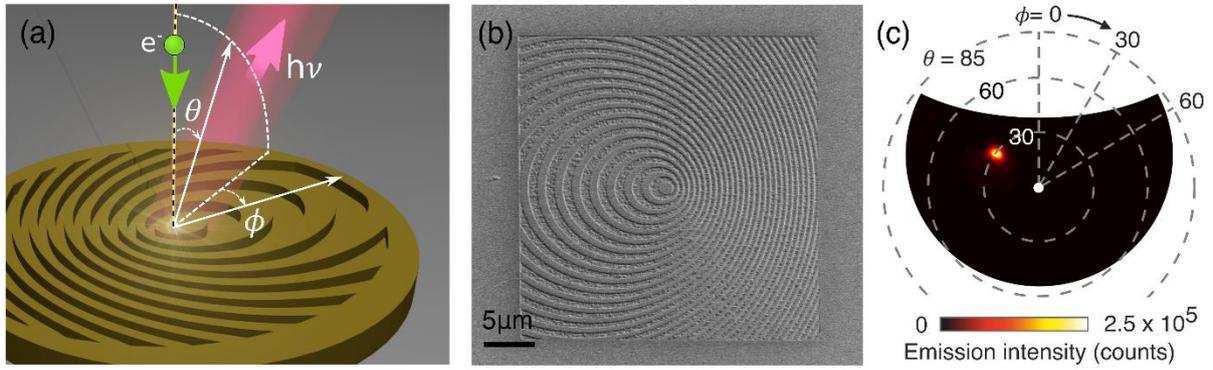

**Figure 1.** (a) Schematic illustration of a free-electron holographic light source: The surface-relief pattern is engineered to couple the electromagnetic excitation resulting from normally incident free-electron impact at the central target point to an output beam of a chosen wavelength and wavefront profile in particular polar $\theta$ and azimuthal $\varphi$ directions. (b) Scanning electron microscope image of a gold holographic source [after Ref. 26] designed to produce an output beam at a wavelength of 800 nm at $\theta = 30°$. (c) Angular distribution of $800 \pm 20$ nm light emission induced by 30 keV electron-beam impact at the target point of the holographic source shown in panel (b).

Here we consider and experimentally study holographic control of EIRE from a variety of dielectric and semiconductor (i.e. non-plasmonic) target materials, specifically silica and sapphire – VIS/NIR transparent, relatively low refractive index dielectrics[34, 35] that present strong intrinsic luminescence, silicon – an elemental semiconductor with a relatively high VIS/NIR refractive index,[36] and polycrystalline germanium antimony telluride (GST) – a high-index chalcogenide alloy (best known as a phase-change medium in the context of optical data storage and non-volatile nanophotonic switching[37-39]). A comparison among these materials and prior studies on gold shows that while surface-relief nanostructuring exerts strong control over the coupling of SPPs to propagating free space light modes, it can also offer some level of control over TR, but has no discernible effect on the spatial distribution of incoherent luminescent emission.

The surface-relief nanostructural patterns required to generate a given output beam are obtained (as described in Ref. 26 and summarized in Supplementary Information) as the interference pattern between a 'reference' electromagnetic field generated by the impact of incident electrons and that of the desired 'object' beam. The cylindrically symmetric toroidal distribution of TR can be calculated analytically[33, 40] but for holographic source design purposes is preferably obtained numerically via a 3D finite-element model comprising an oscillating dipole aligned with the surface normal and positioned a short distance $h = 50$ nm ($\ll \lambda$, where $\lambda$ is the wavelength of light) above the surface.[32, 33, 41, 42] While still inevitably excluding incoherent luminescent emission generated beneath the target surface, this model accurately reproduces the full electromagnetic near field on both sides of the surface plane, which is excited by impinging electrons, including SPPs where relevant.

To inform the selection of holographic source design wavelengths we first recorded EIRE spectra for the unstructured target media (Fig. 2a). These are obtained using a scanning electron microscope operating in fixed-spot mode with an electron energy of 30 keV. The emitted light is collected by a parabolic mirror located above the sample, (confocal with the incident electron beam, which passes through a small hole in the mirror) and directed, in these preliminary measurements, to a VIS/NIR spectrometer (Horiba iHR320 imaging spectrometer with nitrogen-cooled detector array). For the purposes of mapping angular distributions of light emission at a given wavelength (as in Fig. 1c above and Fig. 3 below) the beam is instead directed to a bandpass-filtered CCD camera configured to image the parabolic mirror surface (see Supplementary Figure S2).

Holographic sources were designed in all cases to generate plane-wave output beams propagating at a polar angle $\theta = 30°$ to the surface normal. These comprise patterns of offset concentric oval rings around the electron beam injection point, with radial dimensions determined by the emission wavelength and refractive index of the target medium (Fig. 2b). Sources were designed for wavelengths of 800 nm (as per the original study of gold holographic emitters[26]) – a low-emission wavelength for silica and sapphire but near-maximum-emission wavelength for silicon and GST; 1000 nm – the wavelength of peak emission from silica and sapphire; and 550 nm – a low-emission wavelength for all four dielectric/semiconductor media.

For each combination of emission wavelength and target medium the patterns obtained by interference of the computed surface-plane (reference) and desired output (object) fields were converted to binary masks[43, 44] for ease of fabrication by focused ion beam (FIB) milling. Designs were milled to a depth of 60 nm over 20 μm radius circular domain in all cases. They were milled directly into the silicon and GST samples (respectively, a piece of ~500 μm thick double-polished wafer, and a 500 nm thick sputtered and thermally annealed film of $Ge_2Sb_2Te_5$ on a 200 μm polycrystalline Si substrate). The sapphire and silica samples (~250 μm and ~500 μm thick double-polished wafers, respectively) were first selectively coated over the target area with a 10 nm layer of platinum (via electron beam-induced deposition from a gaseous precursor within the FIB milling system) to prevent the local build-up of charge under the ion beam. Patterns were then milled into the underlying dielectric through this layer, which was subsequently removed.

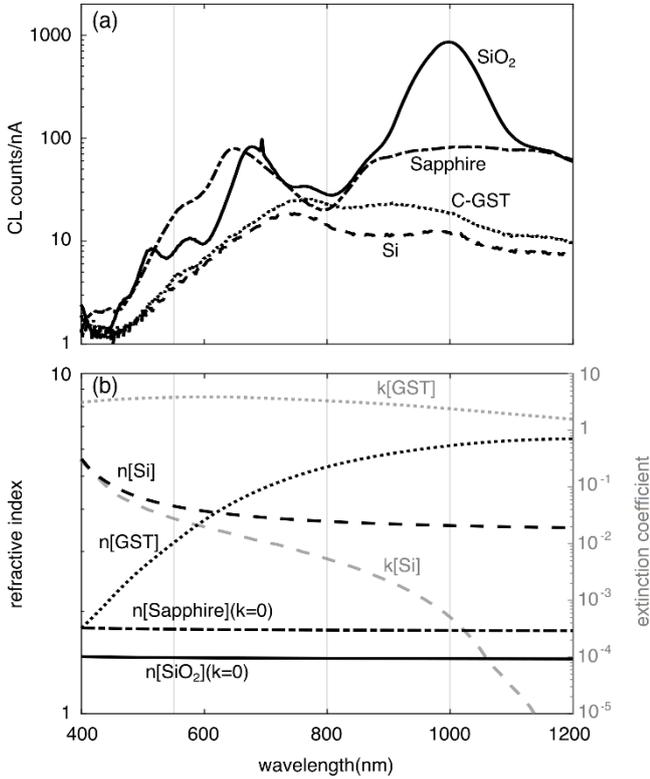

**Figure 2.** (a) Electron-induced light emission intensity spectra, in counts per nA of 30 kV electron beam current, for unstructured polycrystalline GST, silicon, silica and sapphire [as labelled]. (b) VIS-NIR spectral dispersion of the real $n$ and imaginary $k$ parts of refractive index for the same four materials, as used in computational design of holographic sources [data for GST is ellipsometrically measured for the experimental thin film and taken for other materials from Refs. 34-36]. Vertical lines running across panels (a) and (b) denote wavelengths selected for holographic emitter design.

The angular distribution of light emission at each of the design wavelengths (±20 nm) was recorded for each material (Fig. 3), with an electron energy of 30 keV and beam diameter of ~50 nm. Beam current and integration time were adjusted according to target material conductivity and emission brightness, i.e. to avoid sample charging and detector saturation: for GST and silicon an integration time of 60 s was used, with beam currents of 7.5 and 8.5 nA respectively; for silica and sapphire, a time of 8 s and currents of ~6.5 and 1.5 nA. For reference, corresponding emission distributions were also recorded at each wavelength for unstructured regions of each target material. A figure of merit (FOM) for the proportion of light directed by the holographic structure into the intended directional output beam is evaluated as the difference between the fraction of total counts (integrated over the entire emission map) falling within the 'beam spot', which is taken to comprise the brightest pixel within ±20° in $\theta$ or $\varphi$ of the expected output beam direction plus the surrounding pixels with greater than half of that brightness level, and the same fraction evaluated over the same pixels for the reference (unstructured material) emission map. (An ideal device directing all light in a direction to which there is no emission from an unstructured surface of the same material would have a FOM of 1; the gold holographic source of Fig. 1 has a FOM of 0.1502).

In the case of the low-index dielectrics silica and sapphire, the distribution of emitted light from structured surfaces is indistinguishable from that of the unstructured material, which is to say that the holographic patterns provide no discernible control over emission – the FOM at all wavelengths is no higher than the noise level. The spectral dispersion of these materials' electric permittivities is essentially flat over the VIS-NIR range, implying that the same is true of their TR emission. In the spectra of Fig. 2 the TR contribution may thus be taken as the low (short-wavelength) baseline emission level, i.e. as almost negligible against the strength of the intrinsic incoherent luminescence component of emission. This is particularly bright (in terms of photons per electron) for silica at 1000 nm. It then follow that the holographic structures exert no influence over the angular distribution of luminescent emission.

For GST a directional output beam is clearly visible at all three design wavelengths, and for silicon a beam can be discerned at 550 and 800 nm, though in all cases the FOM is at least an order or magnitude lower that of the gold holographic source of Ref. 26. Both materials have rather higher refractive indices than silica and sapphire (though losses are also much higher), and in the case of GST index increases strongly with wavelength. (GST is technically plasmonic at 550 nm, in that it has a negative value of the real part of relative permittivity, however losses are high – SPP propagation length is only ~2 μm) In the knowledge that the holographic structures exert no control over the angular distribution of incoherent luminescent emission and that neither silicon, nor crystalline GST at wavelengths above 620 nm, support SPP propagation, we conclude that the directional beams are derived from transition radiation.

It is clear, not least from the prior study of holographic sources on gold, that the surface-relief structures very effectively couple SPPs – the dominant component of EIRE in that case, to a specified free-space output beam. This is to be expected given their nature as electromagnetic waves bound to the metal/vacuum interface, which can only couple to light in free-space via a scattering structure such as a grating. The ability of the holographic structures to exert an observable level of control over the TR component of coherent emission may be understood to result from the point-like nature of the TR source, whereby light is emitted from the electron injection point (i.e. the singular excitation point around which the holographic structure is designed) with a characteristic lobed 'dipole-above-a-surface' spatial distribution (see Refs. 32, 33, 41, 42 and Supplementary

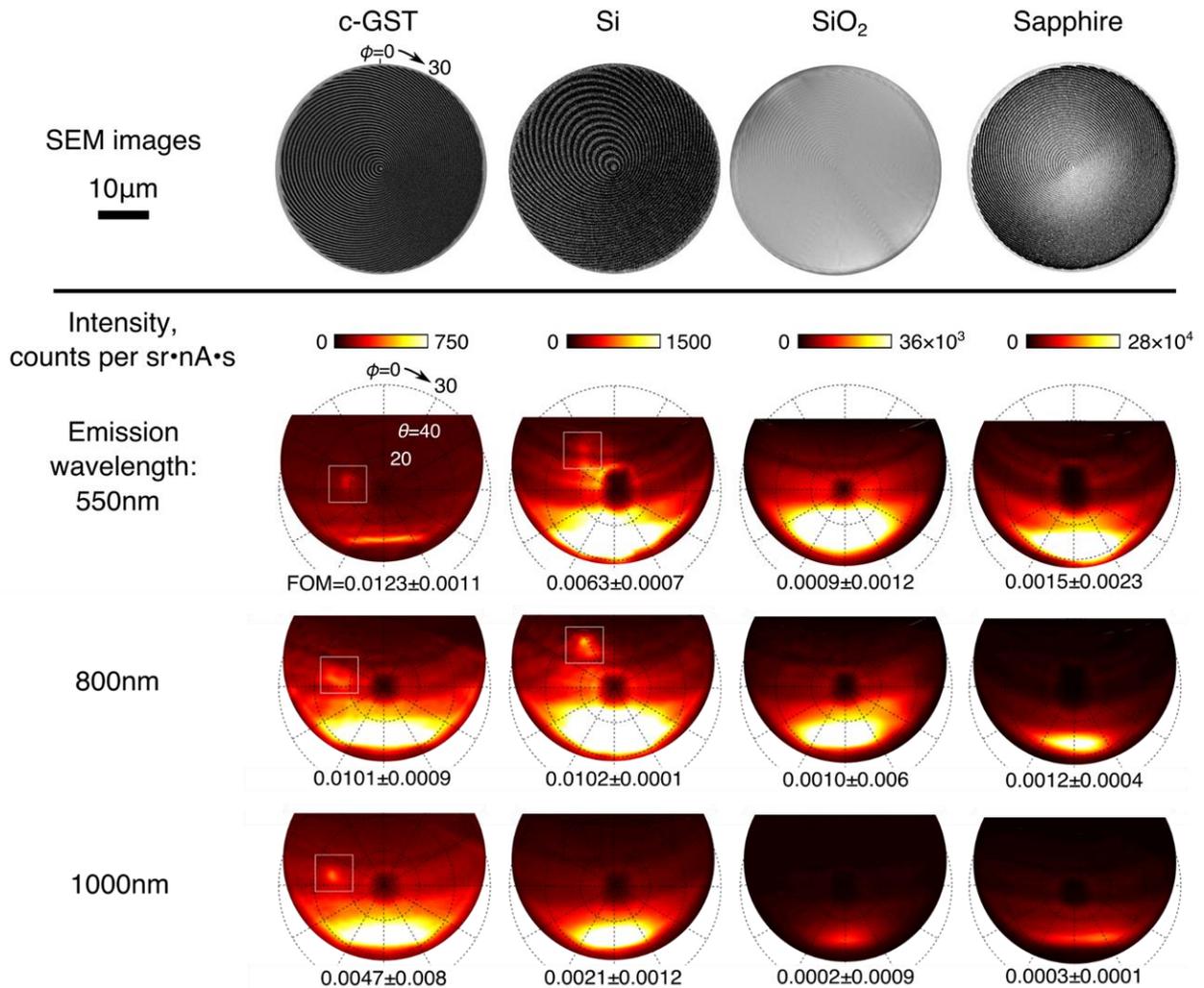

**Figure 3.** Top row: Representative scanning electron microscope images of holographic emitters for each of the four target materials, polycrystalline GST, silicon, silica and sapphire [columns as labelled; variations in imaging contrast/resolution among these reflect variations in electrical conductivity]. Subsequent rows: Angular distribution of electron-beam-induced light emission at 550, 800, and 1000 ± 20 nm [rows as labelled] from holographic surface-relief structures designed for said wavelengths on each target material, with corresponding figures of merit for the proportion of light emitted in the intended $\theta = 30°$ direction. [Azimuthal emission angle $\varphi$ is determined simply by the in-plane orientation of the samples' mirror symmetry axes, and was set to ~300° in all cases. The bright feature at the bottom edge of each emission map is an artefact of mirror geometry/alignment and may be disregarded.]

Fig. S1), such that photons emitted at grazing angles ($\theta \rightarrow 90°$) will scatter from the holographic grating elements to the intended output beam. In contrast, incoherent luminescence is not generated singularly at the electron injection point: it emerges from an interaction volume at least a few microns in diameter, with a Lambertian spatial distribution (a cosine-dependence of emission intensity on polar angle derived from Snell's Law[32]). As such, of the few photons that do emerge at grazing angles, few will do so with an in-plane wavevector matched to the holographic reference field, i.e. they will not scatter to the intended output beam. Indeed, it is found that the output coupling efficiency of such sources decays rapidly as the photon emission (or SPP generation) point is displaced from the designed (electron injection) target point – by a factor $e^{-1}$ within ≤2 μm.[45]

In summary, we have demonstrated that holographically nanostructured surfaces can be engaged to manipulate the spatial distribution of transition radiation (TR) generated by electron beam impact on dielectric/semiconductor surfaces. Surface-relief patterns can be engineered to produce directional output beams at chosen wavelengths, and is most effective (and/or most clearly resolved) for high-refractive-index media in the absence of strong incoherent luminescent emission (which is unperturbed by the holographic structure) and strong plasmonic emission (which can overwhelm the TR signal for metallic target media). The concept offers a means of discriminating between TR and luminescent components of electron-induced light emission in materials analysis and of controlling the output of TR-based electron-beam-driven coherent light sources, such as have been reported in the terahertz and x-ray domains.[46-48]

This work was supported by the Engineering and Physical Sciences Research Council, UK [Project EP/M009122/1], and the Singapore Ministry of Education [grant MOE2011-T3-1-005]. Following a period of embargo, the data from this paper can be obtained from the University of Southampton research